\documentstyle[prl,aps,epsfig,floats]{revtex}

\begin{document}
\title{Quantum Impurities and the Neutron Resonance Peak in ${\bf YBa_2 Cu_3 O_7}$:
Ni versus Zn}
\author{Y. ~Sidis$^1$, P.~Bourges$^1$, H. F. Fong$^2$, B. Keimer$^{2,3}$, L. P. Regnault$^4$, J. Bossy$^5$, A.Ivanov$^6$, B.~Hennion$^1$,\\
P. ~Gautier-Picard$^1$, G.~Collin$^1$, D. L. ~Millius$^7$ and I. A. ~Aksay$^7
$ \\
}
\address{$^1$ Laboratoire L\'eon Brillouin, CEA-CNRS, CE-Saclay, 91191 Gif sur
Yvette, France.\\
$^2$ Department of Princeton University, Princeton, NJ 08544, USA.\\
$^3$ Max-Planck-Institut f\"ur Fertk\"orperforschung, 70569 Stuttgart,
Germany.\\
$^4$ CEA Grenoble, D\'epartement de Recherche Fondamentale sur la Mati\`ere
Condens\'ee, \\
38054 Grenoble cedex 9, France.\\
$^5$ CNRS-CRTBT, BP 156, 38042 Grenoble cedex 9, France.\\
$^6$ Institut Laue Langevin, 156X, 38042 Grenoble cedex 9, France.\\
$^7$ Department of Chemical Engineering, Princeton University, Princeton, NJ
08544 USA.\\
}
\date{\today}



\twocolumn[\hsize\textwidth\columnwidth\hsize\csname@twocolumnfalse\endcsname

\maketitle

\begin{abstract}
The influence of magnetic ($S=1$) and nonmagnetic ($S=0$) impurities on the spin dynamics 
of an optimally doped high temperature superconductor is compared in two samples
with almost identical superconducting transition temperatures:
YBa$_2$(Cu$_{0.97}$Ni$_{0.03}$)$_3$O$_7$ (T$_c$=80 K) and 
YBa$_2$(Cu$_{0.99}$Zn$_{0.01}$)$_3$O$_7$ (T$_c$=78 K).
In the Ni-substituted system, the magnetic resonance peak (which is observed at E$_r 
\simeq$40 meV in the pure system) shifts to lower energy with a preserved E$_r$/T$_c$ ratio 
while the shift is much smaller upon Zn substitution. By contrast Zn, but not Ni, restores 
significant spin fluctuations around 40 meV in the normal state.
These observations are discussed in the light of models proposed for the magnetic resonance 
peak.
 \end{abstract}


]

\narrowtext

Magnetic and nonmagnetic impurities have long been exploited to elucidate
the microscopic nature of the superconducting state. In the cuprate
superconductors, substitution by divalent transition metals for copper
offers a particularly attractive way of introducing such impurities, as
these preserve the doping level and introduce only minimal structural
disorder. The effects of these impurities on the superconducting properties
are unusual and in many ways opposite to those observed in conventional
superconductors. In particular, nonmagnetic Zn$^{2+}$ ions (3d$^{10}$, S=0)
induce a T$_c$ reduction ($\sim$ -12 K/$\%$ Zn) almost three times larger
than magnetic Ni$^{2+}$ ions (3d$^8$, S=1) \cite
{Tarascon88,Mendels99}. Strong-correlation models attribute the large effects of Zn
impurities to a disruption of either local antiferromagnetic \cite
{poilblanc94} or resonating-valence-bond \cite{kilian99} correlations by
spin vacancies, which can lead to a spatially extended bound state. Magnetic
Ni impurities, on the other hand, are coupled to their environment through
exchange interactions and hence act as a much weaker, local scattering
center. Alternative models based on a local charge imbalance induced by
differences in the hybridization of the transition metal ion with oxygen
ligands have also been proposed to explain the disparate effects of Zn and
Ni, without expressly invoking strong correlations \cite{gupta99}.

Here we explore the interplay between both types of impurity and collective
spin excitations in the YBa$_2$Cu$_3$O$_7$ system by inelastic neutron
scattering (INS).
In pure YBa$_2$%
Cu$_3$O$_7$, the spin excitation spectrum is dominated by a magnetic
resonance peak \cite{Rossat91,Mook93,Fong95,Bourges96,Fong96} that is
located at the antiferromagnetic (AF) wave vector {\bf Q}$_{AF}$=($%
\pi/a,\pi/a$) and energy E$_r$$\simeq$40 meV and disappears above T$_c$. The
recent discovery of a similar magnetic resonance peak in the superconducting
state of Bi$_2$Sr$_2$CaCu$_2$O$_{8+\delta}$ \cite{Bi2212} demonstrates that
this collective spin excitation is a generic feature of the cuprates.
There have been many attempts to explain this unusual mode
theoretically and link it to the mechanism of high temperature
superconductivity. Through impurity substitutions, it is possible to vary
the superconducting transition temperature without changing the carrier
concentration, so that the influence of both parameters on the resonance
peak can be separated. However, the first step in an attempt
to follow this program in lightly Zn-substituted YBa$_2$Cu$_3$O$_7$ 
yielded entirely unexpected results \cite{YBCO-Zn,Sidis96}.
In particular, spin excitations around E$_r$ were found to increase
in intensity and persist above ${\rm T_c}$. In order to find out which of
these findings are general consequences of disorder in the ${\rm CuO_2}$
planes and which can be ascribed to the purported resonant scattering of
carriers from Zn impurities, we have undertaken an analogous investigation
of Ni-substituted YBa$_2$Cu$_3$O$_7$.

We present a comparative study of spin excitation spectra in two samples
that were chosen because of their closely similar transition temperatures:
YBa$_2$(Cu$_{0.97}$Ni$_{0.03}$)$_3$O$_7$ (T$_c$=80 K) and YBa$_2$(Cu$_{0.99}$%
Zn$_{0.01}$)$_3$O$_7$ (T$_c$=78 K). Both crystals were grown using the top
seed melt texturing method
and had volumes of $\sim$2 cm$%
^3$. The samples were heat-treated to achieve full oxygenation \cite{YBCO-Zn}%
, and the Ni/Cu and Zn/Cu ratios were deduced from the reduction of T$_c$ as
compared to the pure system. The Ni content and homogeneity was
cross-checked by microprobe measurements.
INS experiments on Ni and Zn substituted samples
have been performed on the triple axis spectrometers 2T at the Laboratoire
L\'eon Brillouin, Saclay, and IN8 at the Institut Laue Langevin, Grenoble
(France).
Focussing Cu(110) monochromator and  PG(002) analyzer were used and a
pyrolytic graphite filter inserted into the scattered beam in order to
remove higher order contamination. The data were taken with a fixed final
energy of 35 meV, and with the crystals in two different orientations where
wave vector transfers of the form ${\bf Q}=(H,H,L)$ and $(3H,H,L)$,
respectively, were accessible. Throughout this article, the wave vector {\bf %
Q} is indexed in units of the reciprocal tetragonal lattice vectors $2\pi
/a=2\pi /b=1.63$ \AA$^{-1}$ and $2\pi /c=0.53$ \AA$^{-1}$. In this notation
the $(\pi/a,\pi/a)$ wave vector parallel to the ${\rm Cu O_2}$ planes
corresponds to points of the form (h/2,k/2) with h and k odd integers.
Because of the well known intensity modulation of the low energy spin
excitations due to interlayer interactions \cite{Rossat91,Mook93,Fong95,Bourges96,Fong96}, 
the data were taken at $L=1.7 l$
where $l$ is an odd integer.

\begin{figure}[t]
\epsfxsize=6cm
$$
\epsfbox{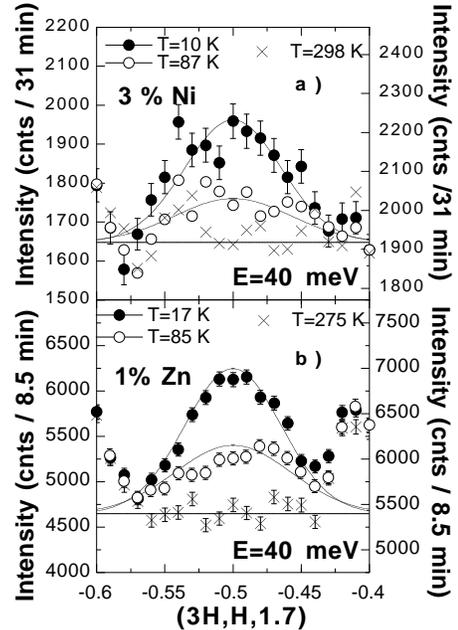}
$$
\caption{ Constant energy scans performed at 40 meV around {\bf Q}$_{AF}$%
=(-1.5,-0.5,1.7) along the (3,1,0) direction: b) YBa$_2$(Cu$_{0.97}$Ni$%
_{0.03}$)$_3$O$_7$, b) YBa$_2$(Cu$_{0.99}$Zn$_{0.01}$)$_3$O$_7$. The right
scale corresponds to data close to room temperature. The peaks
have been fitted to Gaussians.
}
\label{fig1}
\end{figure}

Fig.~\ref{fig1}.a shows a typical constant-energy scan for YBa$_2$(Cu$_{0.97}
$Ni$_{0.03}$)$_3$O$_7$ which reveals that, at temperatures below T$_c$=80 K,
a magnetic signal is present and peaked at {\bf Q}$_{AF}$. With increasing
temperature, the magnetic intensity diminishes drastically, and above T$_c$
only an upper limit corresponding to about 1/3 of the low temperature
intensity can be established, as in pure YBa$_2$Cu$_3$O$_7$ 
\cite{Bourges96,Fong96}. A series of
constant-energy scans at low temperature shows that neither the q-width nor
the peak position exhibit any strong energy dependence. The intrinsic
q-width of the magnetic scattering, $\Delta$Q=0.49$\pm$0.07 \AA$^{-1}$, was
extracted by fitting these profiles to Gaussians. It is larger than $\Delta$Q%
$\simeq$0.25\AA$^{-1}$ which characterizes the magnetic resonance peak in
both pure and 0.5$\%$ Zn substituted systems \cite{YBCO-Zn}.

\begin{figure}[t]
\epsfxsize=6cm
$$
\epsfbox{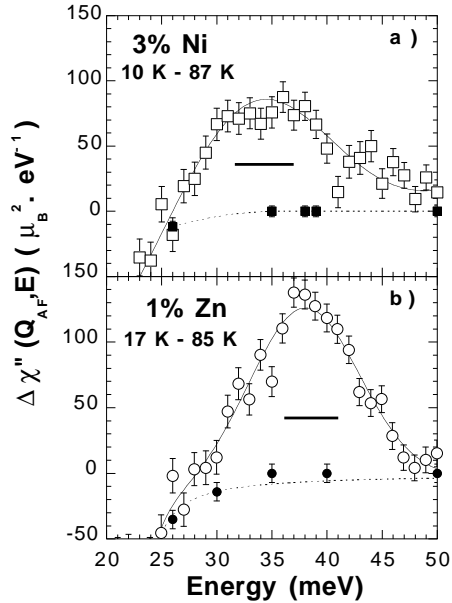}
$$
\caption{Difference between spin excitation spectra, $\protect\chi "$({\bf Q}%
$_{AF}, E$), {\bf Q}$_{AF}$=(-1.5,-0.5,1.7), at low temperature and T$_c$+7 K, calibrated in absolute unit:
a) YBa$_2$(Cu$_{0.99}$Zn$_{0.01}$)$_3$O$_7$ b) YBa$_2$(Cu$_{0.97}$Ni$_{0.03}$%
)$_3$O$_7$. Solid lines are guides to the eye. }
\label{fig2}
\end{figure}

In order to extract the energy dependence of the additional scattering below 
${\rm T_{c}}$, two energy scans have been performed at low temperature ($%
{\rm T\ll T_{c}}$) and at ${\rm T=T_{c}+7K}$, with the wave vector fixed at 
{\bf Q}$_{AF}$. Their difference is reported in Fig.~\ref
{fig2}.a. The extra magnetic intensity in the superconducting state appears
on top of a reference level that becomes slightly negative with decreasing
energy, owing to the thermal enhancement of the nuclear background that
follows the Bose excitation factor. This reference level is determined from
the difference between constant-energy scans performed at low temperature
and at T$_{c}$+7 K (dashed lines and full symbols in Fig.~\ref{fig2}). As
shown in the figure, the enhancement of the magnetic intensity in the
superconducting state is concentrated around a characteristic energy of 35
meV. Deconvolution of the instrumental resolution (width 5 meV) yields an
intrinsic energy width of $\Delta $E$\sim $11 meV.

\begin{figure}[t]
\epsfxsize=6cm
$$
\epsfbox{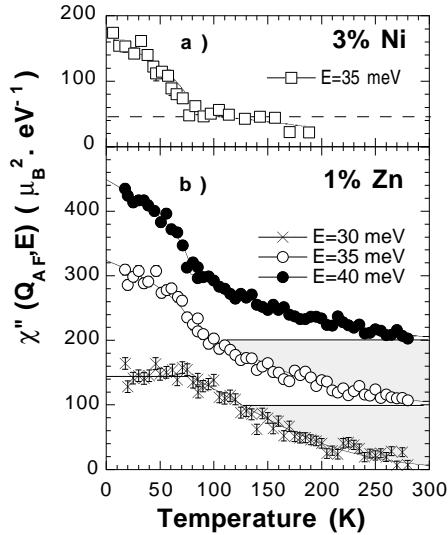}
$$
\caption{Temperature dependence of $\protect\chi "$({\bf Q}$_{AF}, E$): a)
YBa$_2$(Cu$_{0.97}$Ni$_{0.03}$)$_3$O$_7$, E=35 meV. b) YBa$_2$(Cu$_{0.99}$Zn$%
_{0.01}$)$_3$O$_7$, E=30, 35, 40 meV. Data sets are
shifted from one another by 100 $\protect\mu _{B} ^2 eV ^{-1}$. The dashed
line in a) corresponds the upper limit of magnetic response left in the
normal state. }
\label{fig3}
\end{figure}

The magnetic intensity has then been converted to the imaginary part of the
dynamical magnetic susceptibility $\chi"$ after correction by the detailed
balance and magnetic form factors and calibrated against optical phonons
according to a standard procedure \cite{YBCO-Zn}. The temperature dependence
of $\chi"$({\bf Q}$_{AF}$,35 meV), shown in Fig.~\ref{fig3}.a, exhibits a
marked upturn at T$_c$, and an order-parameter-like curve in the
superconducting state: the telltale signature of the resonance peak in the
pure system. The energy-integrated magnetic spectral weight at low
temperature in the energy range probed by the neutron experiment, $\int dE
\chi"({\bf Q}_{AF},E)$, is $1.6 \pm 0.5 \mu_B^2$. Within the errors, this
result is identical to that previously obtained for the resonance peak in
pure YBa$_2$Cu$_3$O$_7$. At {\it fixed} ${\bf Q}={\bf Q}_{AF}$, damping by
the Ni impurities thus leads to a mere redistribution of the spectral weight
of the resonance peak. However, as a consequence of the broadening of the
resonance peak in momentum space the spectral weight obtained after {\it both%
} integrating over energy {\it and} {\bf Q}-averaging over the two
dimensional Brillouin zone ($\sim$0.2 $\mu_B^2$) is almost five times larger
than in pure YBa$_2$Cu$_3$O$_7$ (0.043 $\mu_B^2$). This situation is
closely analogous to pristine, optimally doped Bi$_2$Sr$_2$CaCu$_2$O$%
_{8+\delta}$ (${\rm E_r = 43}$ meV) which is also thought to contain a small
number of intrinsic defects \cite{Kitaoka94}.

The absolute unit calibration also allowed us to quantify the upper limit on
the normal-state intensity: Within the range of our measurements (20-50
meV), $\chi "$ does not exceed 45 $\mu _{B}^{2}eV^{-1}$ in the normal state.
Note that this upper limit, shown as a dashed line in Fig.~\ref{fig3}.a, has
been pushed below the limit previously established for pure YBa$_{2}$Cu$_{3}$%
O$_{7}$ (70 $\mu _{B}^{2}eV^{-1}$) due to various improvements in the
experiments \cite{Bourges_miami}. All features of the magnetic response in
both
superconducting and 
normal states of YBa$_{2}$(Cu$_{0.97}$Ni$%
_{0.03}$)$_{3}$O$_{7}$ are thus closely analogous to the pure system. The
resonance peak is broadened and shifted down in energy, but the ratio E$_{r}$%
/k$_{B}$T$_{c}\sim $5 is preserved.
Further, 3\% Ni substitution induces no
measurable changes in the normal state spectrum.

The situation is very different for Zn impurities. In order to compare the
effects of Zn and Ni on an equal level, we now discuss neutron measurements
taken on a YBa$_2$(Cu$_{0.99}$Zn$_{0.01}$)$_3$O$_7$ crystal whose T$_c$=78 K
is almost identical to that of YBa$_2$(Cu$_{0.97}$Ni$_{0.03}$)$_3$O$_7$.
Again, the magnetic signal is peaked at {\bf Q}$_{AF}$ (Fig.~\ref{fig1}.b),
but only about half of it is removed upon increasing the temperature up T$_c$%
. Above ${\rm T_c}$, the magnetic response gradually fades away and
completely vanishes at room temperature (Fig. 3). Its intrinsic q-width, $
\Delta$Q=0.44$\pm$0.07 \AA$^{-1}$, is larger than in the pure system (0.25\AA$^{-1}$) but comparable to YBa$_2$(Cu$_{0.97}$Ni$_{0.03}$)$_3$O$_7$.
A difference scan in the energy range E=20-50 meV (analogous to that
reported above for the Ni-substituted system) shows that the enhanced $\chi"$%
({\bf Q}$_{AF},\omega)$ in the superconducting state is concentrated around
a characteristic energy $\sim$38 meV, with an intrinsic energy width $\Delta$%
E $\sim$9 meV (fig.~\ref{fig2}.b).

All of these features are closely analogous to those observed in YBa$_2$(Cu$%
_{0.995}$Zn$_{0.005}$)$_3$O$_7$. It is, in fact, surprising that the energy
widths of the magnetic response in both systems are identical (but much
larger than in the pure system) despite the factor-of-two difference in Zn
concentration. The present data set is, however, more comprehensive and
reveals a subtle but distinct difference between the normal-state and
superconducting-state response functions that had gone unnoticed before.
Specifically, the temperature dependences of $\chi"$({\bf Q}$_{AF}, \,35 \,%
{\rm meV})$ and $\chi"$({\bf Q}$_{AF}, \, 40 \,{\rm meV})$ (close to the
maximum of the difference scan of fig.~\ref{fig2}.b) show inflection points
near ${\rm T_c}$ whereas no such feature is apparent in the temperature
dependence of $\chi"$({\bf Q}$_{AF}, \,30 \,{\rm meV})$ (farther away from
the maximum). Moreover, above ${\rm T_c}$ $\chi"$({\bf Q}$_{AF}, \,30 \,{\rm %
meV})$ is actually larger than $\chi"$({\bf Q}$_{AF}, \,35 \,{\rm meV})$ and 
$\chi"$({\bf Q}$_{AF}, \, 40 \,{\rm meV})$. Taken together, these
observations imply that the characteristic energy of the normal-state
response is somewhat lower than the one in the superconducting state.

We can summarize the experimental observations as follows. While the width
of the resonance peak is very sensitive to both types of impurities (and
hence presumably to any type of disorder), there are also pronounced
differences in the magnetic response functions of lightly Zn- and
Ni-substituted ${\rm YBa_2 Cu_3 O_7}$ with identical ${\rm T_c}$. Whereas Ni
impurities do not measurably enhance the normal-state response, a broad peak
with characteristic energy comparable to (but somewhat lower than) the
energy of the resonance peak in pure ${\rm YBa_2 Cu_3 O_7}$ appears in the
normal-state response of Zn-substituted systems. In the superconducting
state, on the other hand, the impurity-induced shift of the magnetic
response is larger in the Ni-substituted system.

Theoretical work on the interplay between collective spin excitations and
quantum impurities in high temperature superconductors is just beginning.
Bulut  \cite{bulut99}
has introduced a model that attributes the normal-state peak in
Zn-substituted systems to impurity-induced Umklapp scattering. In the
framework of this model, the weaker modification of the normal-state
response by Ni could simply be a consequence of the weaker scattering cross
section of Ni. In order to explain the stronger shift of the resonance peak
due to Ni impurities in the superconducting state, one may have to resort to
models that also address the microscopic mechanism underlying the large
difference in scattering cross sections. On a phenomenological level, our
data are consistent with a scenario in which Zn impurities are surrounded by
extended regions whose magnetic properties are strongly modified already far
above ${\rm T_c}$, and in which superconductivity never develops \cite{Sidis96};
superconductivity is then confined to (perhaps only rather narrow) regions
far from the Zn impurities. This would explain why Zn impurities all but
eradicate the effect of superconductivity on the spin excitations which is
so readily apparent in the pure system.

In YBa$_2$(Cu$_{0.97}$Ni$_{0.03}$)$_3$O$_7$, on the other hand, there is no
indication that impurities disrupt the spin correlations over an extended
range. Rather, our data suggest that the magnetic response is not
essentially different from that of a homogeneous system. This scenario is
also consistent with NMR experiments. While these have uncovered a variety
of unusual effects in Zn-substituted ${\rm YBa_2 Cu_3 O_7}$ (such as at
least two different $^{63}$Cu relaxation times that have been attributed to
copper sites at different distances to the Zn impurities \cite{Ishida93}, or
local moments on copper sites adjacent to Zn \cite{Mahajan94}), no such
effects have been reported for Ni-substituted systems.

If, as both neutron and NMR experiments indicate, the spin dynamics of ${\rm %
YBa_{2}(Cu_{1-x}Ni_{x})_{3}O_{7}}$ is indeed amenable to an ``effective
medium'' description, our measurements can be directly compared to current
models of the spin excitations in the cuprates, most of which do not
explicitly incorporate disorder. Specifically, recent theoretical interpretations
of the resonance peak fall into two categories, attributing the mode to
collective modes in the particle-particle \cite{Demler98} and
particle-hole (\it{e.g.}\rm\ Refs. \cite{Lee99,Onufrieva99})
channels, respectively. In
the former model, the magnetic resonance peak is identified with the
so-called $\pi$-excitation that can be visualized as a spin triplet pair of
electrons with center of mass momentum ($\pi$/a,$\pi/a$). The sharp resonance
observed in INS measurements can be ascribed to resonant scattering of
Cooper pairs into $\pi$-pairs \cite{Demler98}. In this
model, the intensity of the magnetic resonance peak is controlled by the
magnitude of the $d$-wave order parameter, whereas the mode energy depends
on hole doping only.

The competing model attributes the magnetic resonance peak to a magnon-like bound state that is pulled below the continuum of spin flip
excitations by antiferromagnetic interactions \cite
{Onufrieva99}. As this resonant excitation is a consequence of
a delicate interplay between the superconducting gap, the shape of the Fermi
surface in the normal state, and the antiferromagnetic spin correlations,
its energy should be very sensitive to the magnitude of the gap. At least in
the optimally doped and overdoped regimes of the phase diagram, where
``pseudogap" effects are not important, the gap is expected to scale with 
${\rm T_c}$.
The shift of the magnetic resonance peak upon Ni substitution, which reduces 
${\rm T_c}$ without changing the hole content, therefore appears
inconsistent with a description in terms of the $\pi$-mode. In contrast, the
fact that the ratio E$_r$/T$_c$ remains constant suggests that the
collective mode and the superconducting gap are renormalized in the same
way, which is consistent with the spin-exciton scenario for the magnetic
resonance peak \cite{Onufrieva99}. Of course, a direct
measurement of the superconducting gap is required to confirm this
interpretation. Finally, we note that the impurity broadening of magnon-like excitations in $d$-wave superconductors predicted using general arguments
\cite{vojta99} is in good overall agreement with the widths of the resonance
peaks reported here. 

In conclusion, our INS data show that the spin excitation spectra of YBa$_2$%
(Cu$_{0.97}$Ni$_{0.03}$)$_3$O$_7$ and YBa$_2$(Cu$_{0.99}$Zn$_{0.01}$)$_3$O$_7
$ are still dominated by a magnetic resonance peak below T$_c$, but that the
two types of impurity affect the spin correlations in very different ways.
The interplay between quantum impurities and collective spin dynamics in the
cuprates is therefore a surprisingly rich new field of investigation.

The work at Princeton University was supported by the NSF MRSEC Program under
DMR-9809483.

\end{document}